\documentclass[conference]{IEEEtran}
\usepackage{cite}

\ifCLASSINFOpdf
\else
  \usepackage[dvips]{graphicx}
  \graphicspath{{../eps/}}
  \DeclareGraphicsExtensions{.eps}
\fi
\usepackage{mathrsfs}
\usepackage{amssymb}
\usepackage[cmex10]{amsmath}
\usepackage{array}

\usepackage{mdwmath}
\usepackage{mdwtab}

\usepackage[font=footnotesize]{subfig}
\usepackage{fixltx2e}

\usepackage{stfloats}

\usepackage{slashbox}
\usepackage{xcolor}

\begin{document}
%
\title{Heterogeneous Recovery Rates against SIS Epidemics in Directed Networks}

\author{\IEEEauthorblockN{Bo Qu}
\IEEEauthorblockA{Delft University of Technology\\
Delft, Netherlands\\
Email: b.qu@tudelft.nl}
\and
\IEEEauthorblockN{Alan Hanjalic}
\IEEEauthorblockA{Delft University of Technology\\
Delft, Netherlands\\
Email: a.hanjalic@tudelft.nl}
\and
\IEEEauthorblockN{Huijuan Wang}
\IEEEauthorblockA{Delft University of Technology\\
Delft, Netherlands\\
Boston University\\
Boston, USA\\
Email: h.wang@tudelft.nl
}}


%


\maketitle

\begin{abstract}
\boldmath
The nodes in communication networks are possibly and most likely equipped with different recovery resources, which allow them to recover from a virus with different rates. In this paper, we aim to understand know how to allocate the limited recovery resources to efficiently prevent the spreading of epidemics. We study the susceptible-infected-susceptible (SIS) epidemic model on directed scale-free networks. In the classic SIS model, a susceptible node can be infected by an infected neighbor with the infection rate $\beta$ and an infected node can be recovered to be susceptible again with the recovery rate $\delta$. In the steady state a fraction $y_\infty$ of nodes are infected, which shows how severely the network is infected. Instead of considering the same recovery rate for all the nodes in the SIS model, we propose to allocate the recovery rate $\delta_i$ for node $i$ according to its indegree and outdegree-$\delta_i\scriptsize{\sim}k_{i,in}^{\alpha_{in}}k_{i,out}^{\alpha_{out}}$, given the finite average recovery rate $\langle\delta\rangle$ representing the limited recovery resources over the whole network. We consider directed networks with the same power law indegree and outdegree distribution, but different values of the directionality $\xi$ (the proportion of unidirectional links) and linear correlation coefficients $\rho$ between the indegree and outdegree. We find that, by tuning the two scaling exponents $\alpha_{in}$ and $\alpha_{out}$, we can always reduce the infection fraction $y_\infty$ thus reducing the extent of infections, comparing to the homogeneous recovery rates allocation. Moreover, we can find our optimal strategy via the optimal choice of the exponent $\alpha_{in}$ and $\alpha_{out}$. Our optimal strategy indicates that when the recovery resources are sufficient, more resources should be allocated to the nodes with a larger indegree or outdegree, but when the recovery resource is very limited, only the nodes with a larger outdegree should be equipped with more resources. We also find that our optimal strategy works better when the recovery resources are sufficient but not yet able to make the epidemic die out, and when the indegree outdegree correlation is small.  
\end{abstract}


%
\IEEEpeerreviewmaketitle

\section{Introduction}
Network security has become one of the most important challenges for communication networks.
A network node could be infected, become a new infection source and infect other hosts. On the other hand, network nodes are usually equipped with certain recovery resources, so that they can be recovered from the infection in a finite time. The infection and recovery processes above have been described by epidemic models\cite{albert2000error,kephart1991directed,pastor2001epidemic,garetto2003modeling,ganesh2005effect,canright2006spreading}. One of the most widely studied models is the susceptible-infected-susceptible (SIS) model\cite{anderson1991infectious,bailey1975mathematical,daley2001epidemic,van2011n,cator2013susceptible,li2012susceptible,allen1994some,zhou2003stability,boccara1993critical,shi2008sis,korobeinikov2002lyapunov}. 

In the homogeneous SIS model, at any time $t$, the state of a node is a Bernoulli random variable, where $X_i(t)=0$ represents that node $i$ is susceptible and $X_i(t)=1$ if it is infected. Each infected node infects each of its susceptible neighbors with an infection rate $\beta$. The infection process between any infected node and any susceptible node is an independent Poisson process. The recovery process of each infected node is also an independent Poisson process with a recovery rate $\delta$. In a single network, if the effective infection rate $\tau\triangleq\beta/\delta$ is smaller than a critical value $\tau_c$, i.e. $\tau<\tau_c$, the infection rapidly disappears, whereas if $\tau>\tau_c$, a nonzero fraction $y_{\infty}$ of nodes are being infected in the steady state\cite{anderson1991infectious,pastor2001epidemic,ferreira2012epidemic}. The fraction of infection $y_\infty$, ranging in $[0,1]$, indicates that how severe the influence of the virus is, i.e. the larger the infected fraction $y_\infty$ is, the more severely the network is infected. 

The classic SIS model assumes that the infection rate $\beta$ is the same for all infected-susceptible node pairs and so is the recovery rate $\delta$, thus homogeneity. However, the recovery time of real-world network components can be different, since they can be equipped with, for example, different antivirus softwares and different levels of human intervention that lead to different time to discover and cure the virus. Meanwhile, a fast recovery which requires an antivirus software with a high scanning frequency or frequent human intervention, is costly. The balance of the effect and cost of recovering capacity is of key importance. That is to say, with the limited recovery resources, how to distribute them, is critical. 

In this work, we consider the SIS epidemic spreading with homogeneous infection rate $\beta$ in a network with N nodes. However, we propose heterogeneous recovery rates allocation to minimize the fraction $y_\infty$ of infected nodes in the steady state for the given finite average recovery rate $\langle\delta\rangle=(\sum_{i=1}^{N}\delta_i)/N$ representing the limited recovery resources. 

It has been revealed that in real-world networks such as the Internet\cite{caldarelli2000fractal} and World Wide Web\cite{albert1999internet}, the probability that a node has $k$ connections (degree $k$) follows a scale-free distribution $P(k)\scriptsize{\sim} k^{-\lambda}$\cite{faloutsos1999power,siganos2003power,chen2002origin}, with an exponent $\lambda$ ranging between $2$ and $3$. The studies\cite{ebel2002scale,barabasi2000scale} have also shown the directed characteristics of those networks: when a node $i$ is connected to $j$, $j$ is not necessarily connected to $i$. Hence, we consider directed scale-free (SF) networks which has been less explored for the SIS model. 

We propose to allocate the recovery rate of each node as a function of the indegree and outdegree of the node, i.e. $\delta_i\scriptsize{\sim}k_{i,in}^{\alpha_{in}}k_{i,out}^{\alpha_{out}}$. The degree, which is easily to compute, has been chosen for the following reasons: a) the nodal degree has been shown to be crucial to select e.g. the nodes to immunize to reduce the infection fraction and the nodes to protect to maintain the network connectivity\cite{JCN2013_Trajanovski,huang2013robustness}; b) the degree is correlated with many other network properties of a node\cite{li2012degree,li2011correlation}. When the network is undirected, our strategy becomes $\delta_i\scriptsize{\sim}k_i^{\alpha}$. Our goal is to find the optimal allocation, i.e. the optimal scaling parameter $\alpha_{in}$ and $\alpha_{out}$,  which leads to a minimal infection fraction. We find that, compared to the homogeneous SIS model, where the recovery rate is the same for all the nodes, our optimal strategy can always further evidently reduce the infection fraction $y_\infty$ (even to $0$, thus, the epidemic dies out) in the steady state. The novelty of our approach lies in: a. We generalize the classic homogeneous SIS model to heterogeneous, which allows each node to have a different recovery rate; b. We consider not only the classic undirected networks but also directed networks, characterized by the directionality $\xi$ and the indegree outdegree correlation $\rho$. This introduces heterogeneity into the network structure; c. We propose a heterogeneous recovery rates allocation strategy based on the degree of each node, the simplest network property to compute.  

The main method to reduce the fraction of the infected nodes is immunization. The immunization strategies\cite{cohen2003efficient,madar2004immunization,pastor2002immunization,gomez2006immunization,bai2007immunization} select a given number of nodes to be immunized. When a node gets immunized, it can not get infected nor infect the others. To immunize a node is equivalent to allocating an infinite recovery rate to this node. In this sense, the immunization problem is to decide which set of a given number $m$ of nodes should we assign an infinite recovery rate whereas the rest nodes have the same given and finite recovery rate. Our work addressed the more general case where each node may have a different recovery rate instead of assigning two possible recovery rates to the whole network. Heterogeneous SIS model has been addressed recently \cite{fu2008epidemic,buono2013slow,preciado2013optimal} by taking into account either the heterogeneous infection or recovery rates. \cite{fu2008epidemic} studies the epidemic threshold with variable infection rates. \cite{buono2013slow} also models heterogeneous infection rates and observes slow epidemic extinction phenomenon. \cite{preciado2013optimal} discusses how to choose the infection and recovery rates from given discrete sets to let the virus die out, but our work addresses, more generally, how to reduce the fraction of infected nodes-a zero fraction means the extinction of the virus. 

The rest of this paper is organized as follows. Section \ref{sec2} introduces the directionality $\xi$ and indegree outdegree correlation $\rho$ to describe the directed networks, and the epidemic threshold as well. Section \ref{sec3} illustrates the distribution of the recovery rate $\delta_i$, the infection fraction vs. the epidemic threshold, and the influence of the exponents $
(\alpha_{in}$, $\alpha_{out}$) on the infection fraction, when the recovery rate is distributed heterogeneously according to our strategy. Section \ref{sec4} investigates the optimal exponents $(\alpha_{in,opt}, \alpha_{out,opt})$ for different indegree outdegree correlations $\rho$, and compares the infection fraction obtained by our optimal heterogeneous recovery rates allocation and by homogeneous recovery rates allocation. Section \ref{sec5} concludes the paper.       

\section{Directed networks and the epidemic threshold}
\label{sec2}

\subsection{Directed networks}
In order to describe the directed SF networks, as in our previous works\cite{li2013epidemic,qu2014non}, we employ the directionality $\xi$ and indegree outdegree correlation $\rho$. The directionality is $\xi=L_{unidirectional}/L$, where $L_{unidirectional}$ and $L_{bidirectional}$ represent the number of unidirectional and bidirectional links respectively, and the
normalization $L=L_{unidirectional}+2L_{bidirectional}$, because
a bidirectional link can be considered as two unidirectional links. When $\xi=0$
the network is undirected.

The sum of indegree and the sum of outdegree are equal in a directed
network, but the indegree and outdegree of a single node are
usually not the same. The indegree outdegree correlation $\rho$ is to quantify the possible difference between indegree and outdegree and it is actually the linear correlation coefficient between indegree and outdegree.
\begin{equation}
\rho = \frac{\sum_{i=1}^{N}(k_{i,{in}}-\langle k\rangle)(k_{i,{out}}-\langle k\rangle)}{\sqrt{\sum_{i=1}^{N}(k_{i,{in}}-\langle k\rangle)^2}\sqrt{\sum_{i=1}^{N}(k_{i,{out}}-\langle k
\rangle)^2}}      
\end{equation}
where $k_{i,{in}}$ and $k_{i,{out}}$ are the indegree and
outdegree of node $i$ respectively. The average degree $\langle k
\rangle$ is the same for both indegree and outdegree. Note that when
$\rho=1$ the indegree is linearly dependent on the outdegree for all
nodes, and when $\rho=0$ the indegree and outdegree are independent of
each other. In this paper we confine ourselves to the case in which the
indegree and outdegree follow the same distribution. In this case,
$\rho=1$ implies that $k_{i,{\rm in}}=k_{i,{\rm out}}$ holds for every
node $i$. Particularly, in this paper, we construct the directed SF networks with a given indegree outdegree correlation $\rho$ using the algorithm ANC we proposed in\cite{qu2014non}, so we get the networks with the directionality\footnote{$E(\xi)=1-\langle k\rangle^{2}N/(N-1)^{2},
  \lim_{N\to +\infty}E(\xi)=1$.} $\xi\approx1$ and a specified indegree outdegree correlation $\rho$. 

\subsection{The epidemic threshold of the SIS model}
Given the adjacency matrix\footnote{The entries of the adjacency matrix $A$ are,  $a_{ij}=1$ if node $i$ is connected to node $j$ or otherwise $a_{ij}=0$.} $A$ of a network, the epidemic threshold of the homogeneous SIS model with the same infection rate and recovery rate for all the links and nodes respectively, via the N-intertwined mean field approximation (NIMFA) is $\tau_c=1/\lambda_1(A)$, where $\lambda_1(A)$ is the largest eigenvalue of the adjacency matrix $A$\cite{van2011n}. When the effective infection rate $\tau=\beta/\delta$ is above this threshold $\tau>\tau_c$, the epidemic spreads out. Via the mean field approximation-NIMFA, we can also obtain the sufficient condition for the epidemic to die out in the heterogeneous SIS model, which is $\Re(\lambda_1(-diag(\delta_i)+\beta*A))\leq0$\cite{van2011n}\cite{preciado2013optimal}, where $A$ is the adjacency matrix of the network, $diag(\delta_i)$ is a $N$ by $N$ diagonal matrix with elements $\delta_i$ and $\Re(\lambda_1(\bullet))$ represents the real part of the largest eigenvalue of a matrix\footnote{If $A$ is asymmetric, the largest eigenvalue of the matrix $-diag(\delta_i)+\beta*A$ may be complex.}.  
            
\section{The effect of the heterogeneous recovery rates allocation}
\label{sec3}
Instead of the approximated discrete-time SIS model simulations, we use a continuous-time simulator, proposed by van de Bovenkamp et al. and described in detail in \cite{li2012susceptible}. We are going to evaluate the effect of our degree based heterogeneous recovery rates allocation strategy via the infection fraction $y_\infty$ in the steady state. In this paper, we consider directed SF networks (both the
indegree and outdegree follow the same power law distribution) with the power law exponent $\lambda=2.5$, the networks size $N=1000$, the smallest degree $k_{min}=2$ and the natural degree cutoff $k_{max}=\lfloor N^{1/(\lambda-1)}\rfloor$\cite{PhysRevLett.85.4626}. The directionality $\xi$ and indegree outdegree correlation $\rho$ can be tuned, using the algorithms we proposed in\cite{qu2014non}. For the heterogeneous SIS model, we set the infection rate $\beta=1$ for all links, and assign the recovery rate according to our strategy $\delta_i=c_2k_{i,in}^{\alpha_{in}}k_{i,out}^{\alpha_{out}}$, where the parameter $\alpha_{in}$ and $\alpha_{out}$ can be tuned and the constant number
$c_2$ is decided by the given average recovery rate $\langle\delta\rangle$. All the results are based on the average of $1000$ simulations.

We use the continuous-time simulations which precisely simulate the continuous SIS dynamics, instead of discrete-time ones, because the precision of a discrete time simulation of a continuous process requires a small time bin to sample the continuous process so that within each time bin, no multiple events occur. A heterogeneous SIS model allows different as well large infection or recovery rates, which requires even smaller time bin size and challenges the precision of discrete-time. 

\subsection{The distribution of the recovery rate $\delta_i$}
In this paper, the indegree and outdegree of a node both follow the same power-law distribution
\[
\Pr [D_{in}=k]=\Pr [D_{out}=k]=c_{1}k^{-\lambda }
\]
where $c_{1}=1/{\sum_{k=k_{min}}^{k_{max}}k^{-\lambda }}$. The joint degree distribution of a random node in a directed network with indegree
outdegree correlation $\rho $ and directionality $\xi\approx1$ that we generated
using ANC follows

$\Pr [D_{in}=k_{in},D_{out}=k_{out}]=$
\[\Big\{\begin{tabular}{lc}
$\rho c_{1}k_{in}^{-\lambda }+(1-\rho )c_{1}^{2}k_{in}^{-2\lambda }$, &when 
$k_{in}=k_{out}$ \\ 
$(1-\rho)c_{1}^{2}k_{in}^{-\lambda }k_{out}^{-\lambda }$, &when $k_{in}\neq k_{out}$%
\end{tabular}%
\Big. 
\]

The recovery rate of each node is assigned according to our strategy as $%
\delta _{i}=c_{2}k_{i,in}^{\alpha _{in}}k_{i,out}^{\alpha _{out}}$, where
$c_2 =\frac{\left\langle \delta \right\rangle }{\sum_{k_{in}=1}^{N-1}%
\sum_{k_{out}=1}^{N-1}\Pr [D_{in}=k_{in},D_{out}=k_{out}]
k_{in}^{\alpha _{in}}k_{out}^{\alpha _{out}}}$, so the probability density function of the recovery rate $\Delta$ is 
\[
\Pr [\Delta=x]=\sum_{k_{in}=1}^{N-1}\Pr
[D_{in}=k_{in},D_{out}=\left( \frac{x}{c_{2}k_{in}^{\alpha _{in}}}%
\right) ^{\frac{1}{\alpha _{out}}}].
\]

\begin{figure}[!tb]
\centering
\includegraphics[width=2.8in]{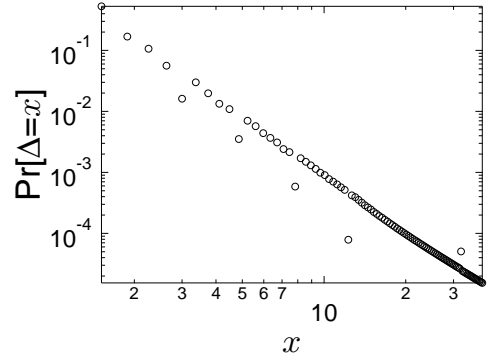}
\caption{The probability distribution of the recovery rate with $\langle\delta\rangle=2$, $\rho=0.5$, $\alpha_{in}=0.3$ and $\alpha_{out}=0.6$.}
\label{fig0}
\end{figure}

In special cases, but not always, the distribution of the recovery rate follows strictly the power law. For example, when the indegree outdegree correlation is $\rho=1$, the distribution follows the power law with the exponent $\alpha_{in}+\alpha_{out}-\lambda$. Fig. \ref{fig0} shows an example of the distribution of the recovery rate with $\langle\delta\rangle=2$, $\rho=0.5$, $\alpha_{in}=0.3$ and $\alpha_{out}=0.6$. The distribution in this case follows approximately a power law.
  
\subsection{The infection fraction $y_\infty$ vs. $\Re(\lambda_1(-diag(\delta_i)+\beta*A))$}

   \begin{figure*}[!tbh]
      \centering
      \subfloat[\label{fig1a}]{%
        \includegraphics[width=2.3in]{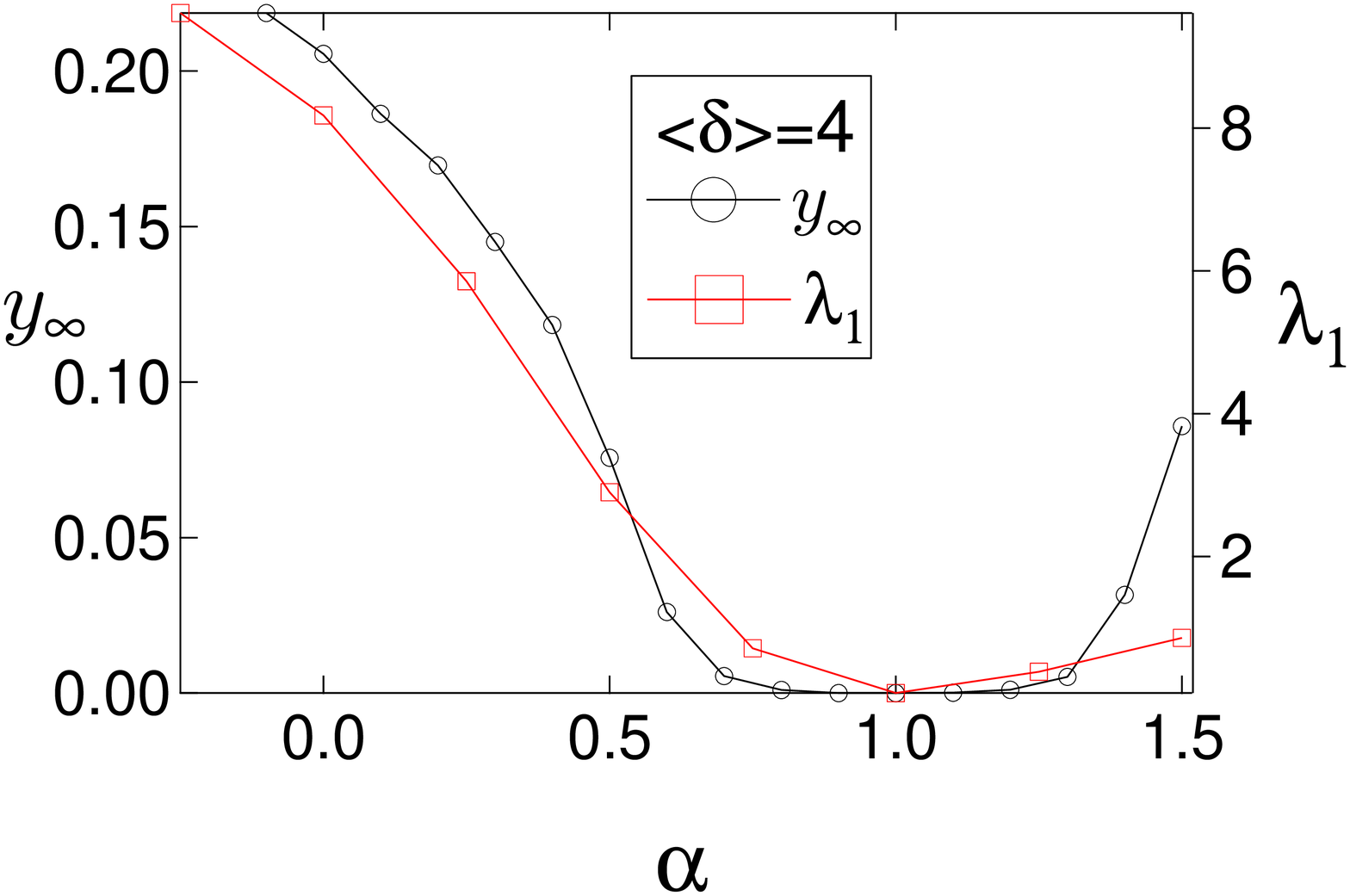}%
      }
      \hfill
      \subfloat[\label{fig1b}]{%
        \includegraphics[width=2.3in]{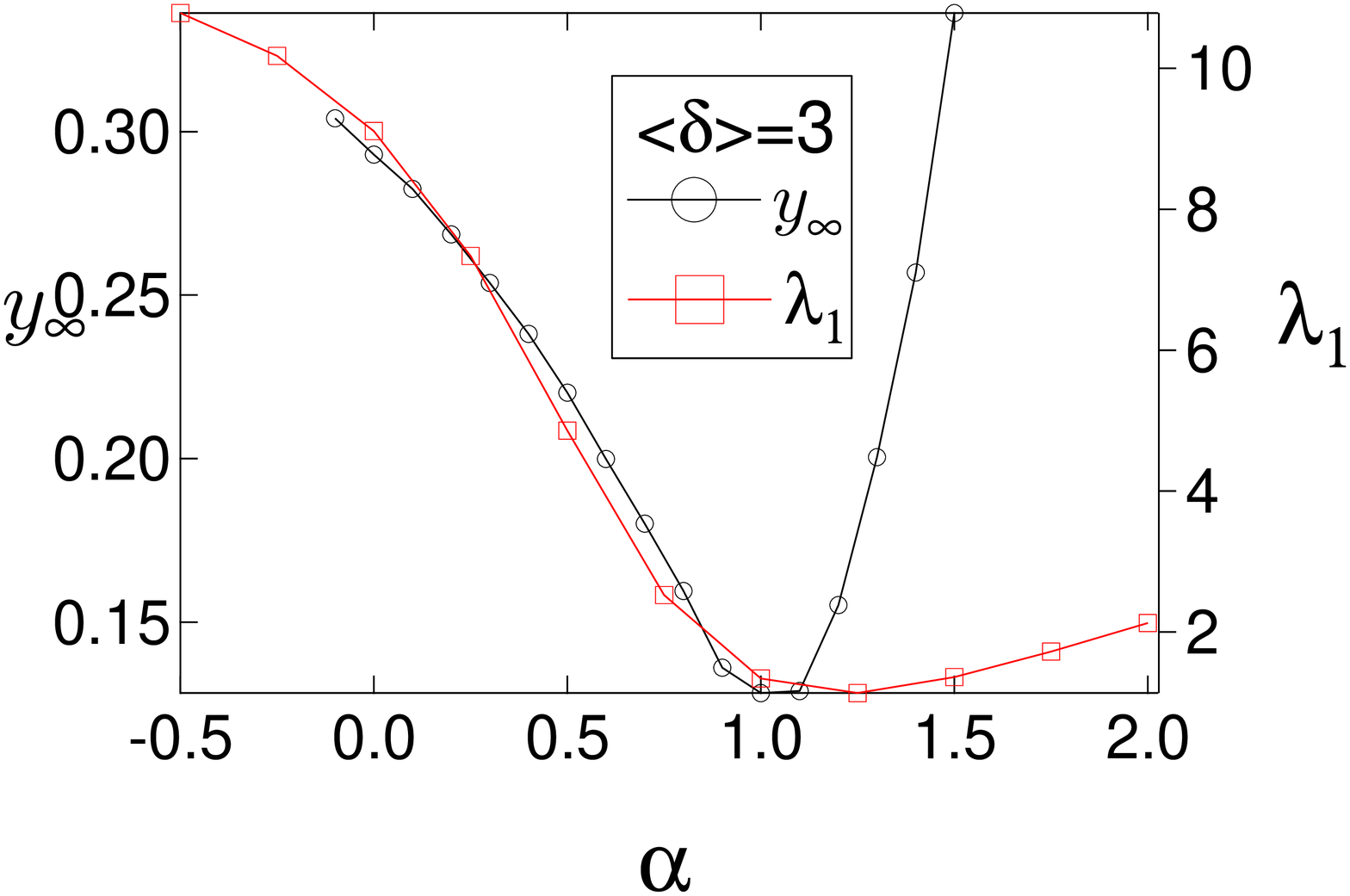}%
      }
      \hfill
            \subfloat[\label{fig1c}]{%
              \includegraphics[width=2.3in]{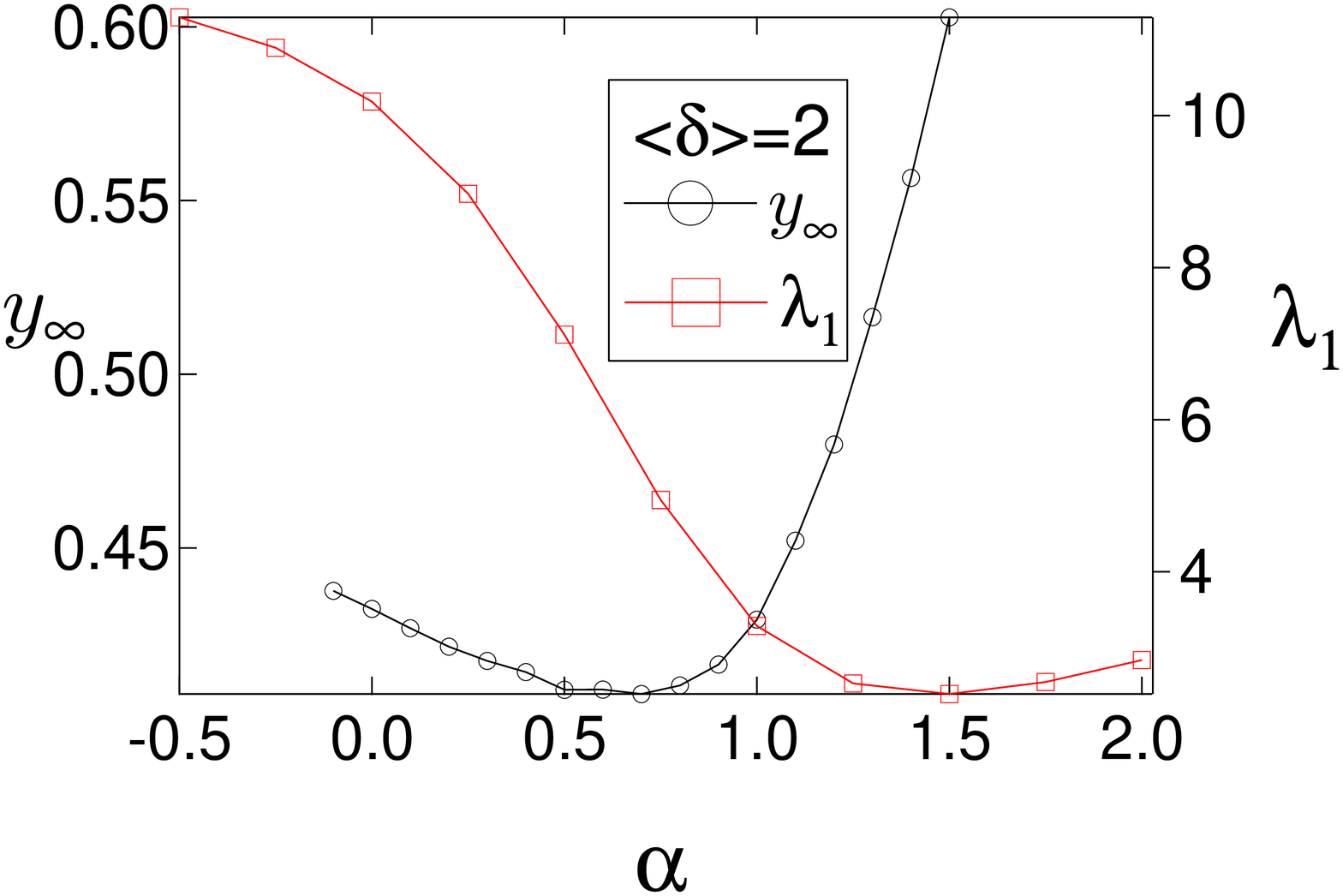}%
            }
      \caption{The plots of the infection infraction $y_\infty$ and the largest eigenvalue of the matrix $-diag(\delta_i)+A$ as a function of the exponent $\alpha$ for different values of the average recovery rate $\langle\delta\rangle$.}
      \label{fig1}
    \end{figure*}
    
           \begin{figure*}[!tb]
              \centering
              \subfloat[$\langle\delta\rangle=3$ and $\rho=0$\label{fig2a}]{%
                \includegraphics[width=2.3in]{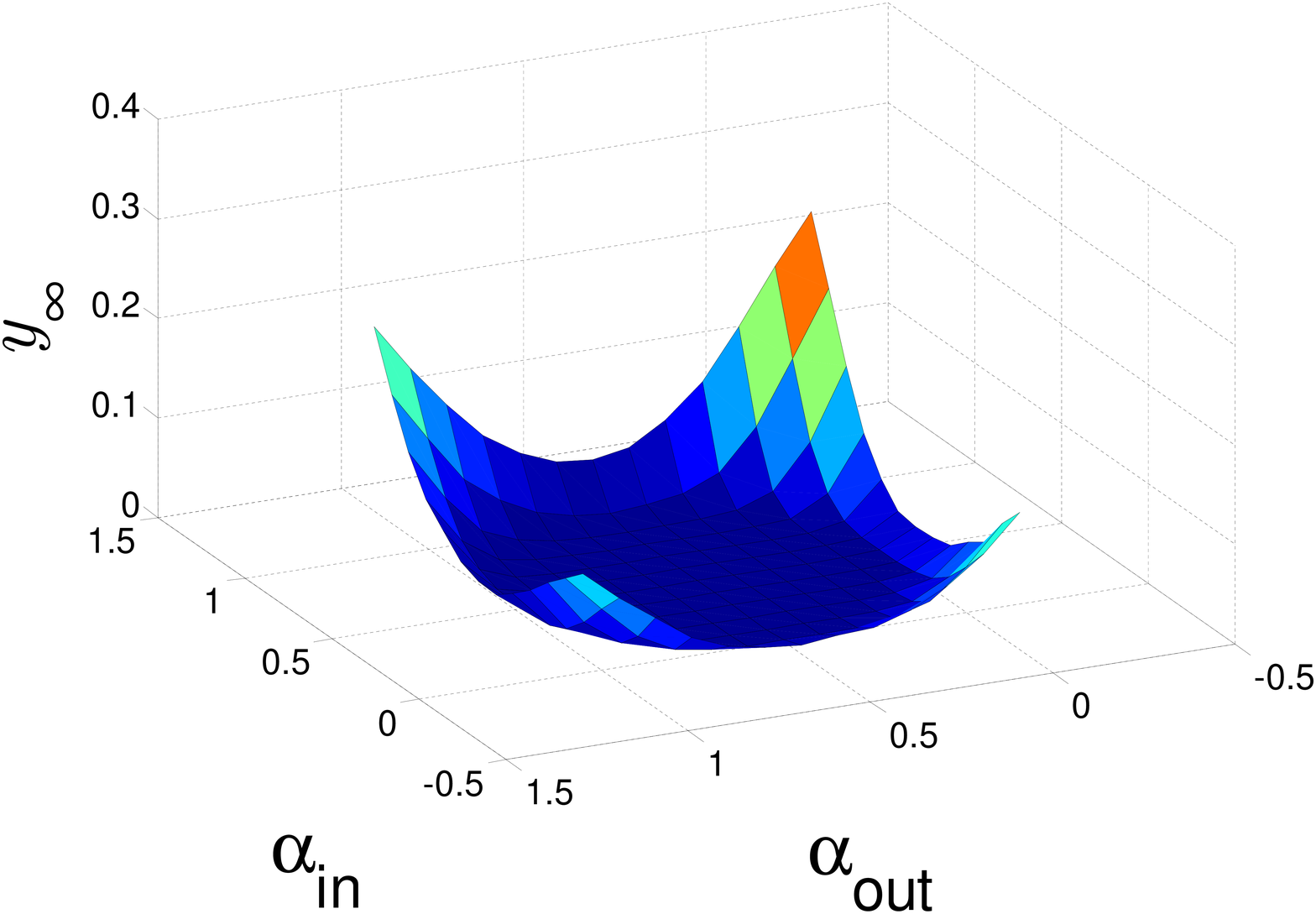}%
              }
              \hfill
              \subfloat[$\langle\delta\rangle=2.5$ and $\rho=0$\label{fig2b}]{%
                \includegraphics[width=2.3in]{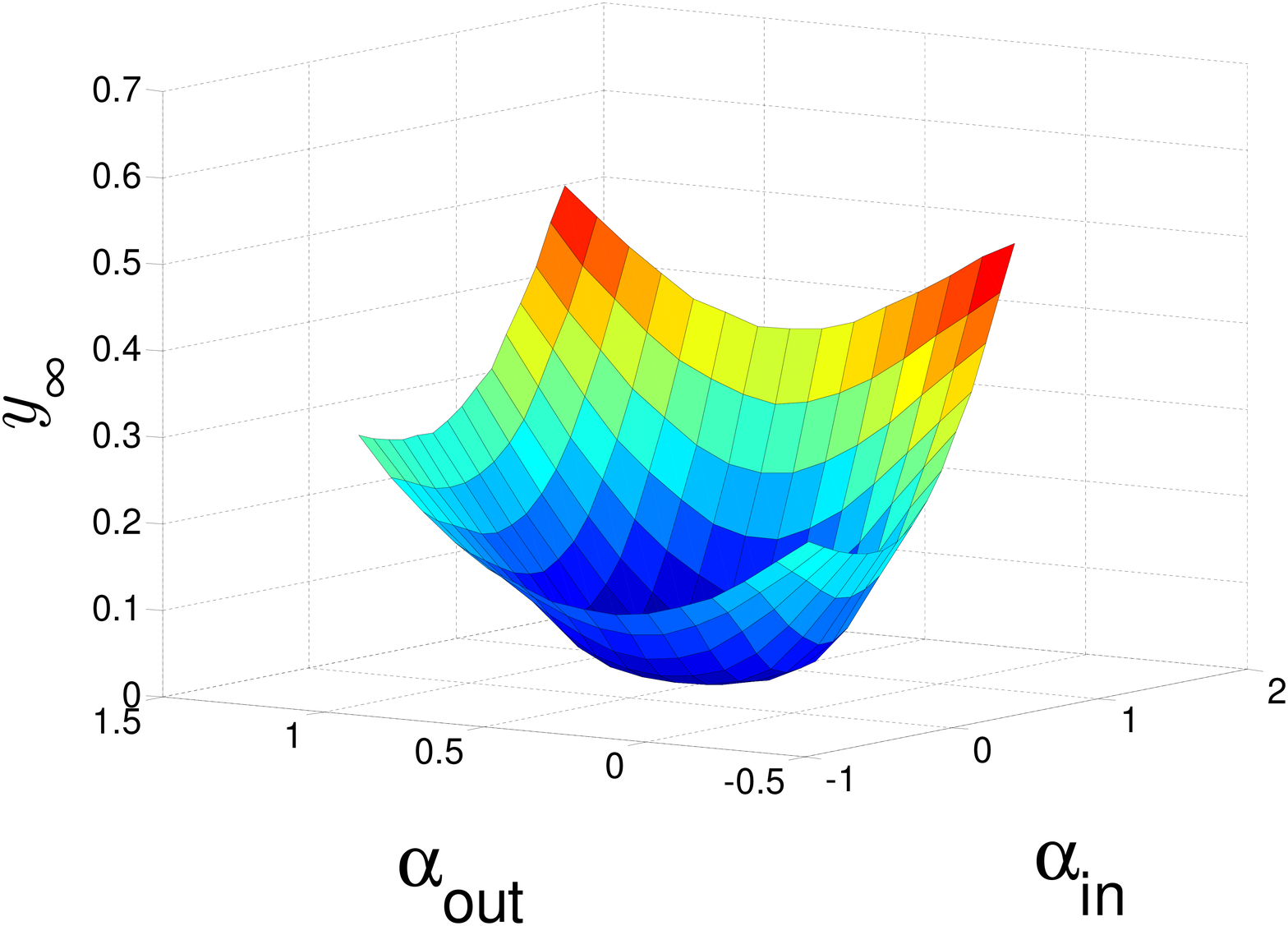}%
              }
              \hfill
                    \subfloat[$\langle\delta\rangle=3$ and $\rho=0.5$\label{fig2c}]{%
                      \includegraphics[width=2.3in]{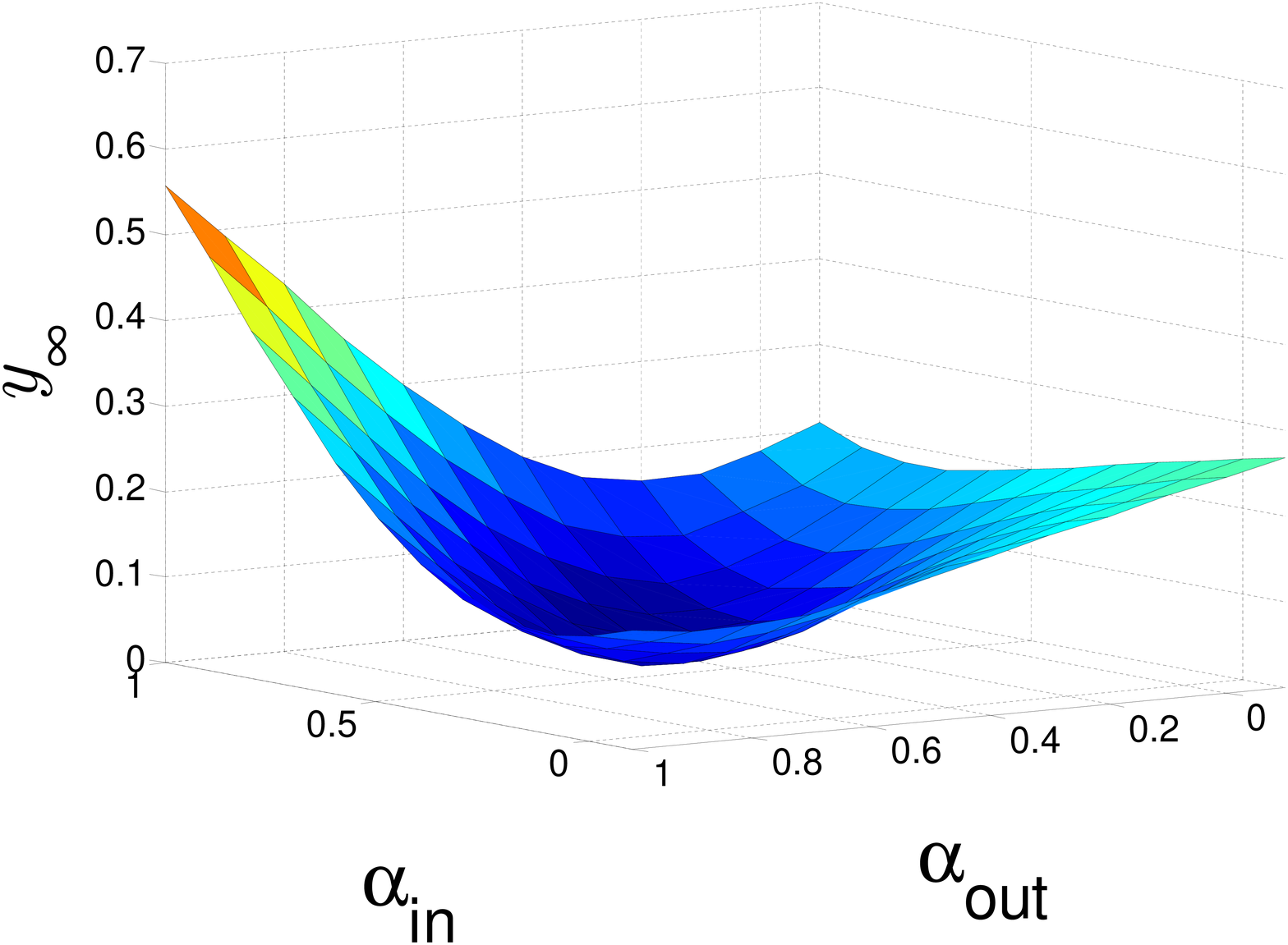}%
                    }
              \caption{The plots of infected infraction $y_\infty$ as a function of the exponents $\alpha_{in}$ and $\alpha_{out}$ with different values of the average recovery rate $\langle\delta\rangle$ and the indegree outdegree correlation $\rho$}
              \label{fig2}
            \end{figure*}
            
From network designers' point of view, we aim to either increase the epidemic threshold or decrease the fraction of infection via network topology modifications, the allocation of the infection rates, recovery rates or immunization\cite{PhysRevE.84.016101,li2012bounds,wang2013effect}. We would like to first understand in the heterogeneous SIS, whether a small $\Re(\lambda_1(-diag(\delta_i)+\beta*A))$ would suggest a low infection fraction. If so, we could simplify the optimization of the fraction of infection to the optimization of the $\Re(\lambda_1(-diag(\delta_i)+\beta*A))$.      
    
We take undirected scale-free networks as an example. In this case, the largest eigenvalue is real, i.e. $\Re(\lambda_1(-diag(\delta_i)+\beta*A))=\lambda_1(-diag(\delta_i)+\beta*A)$. We compare the infection fraction $y_\infty$ in the steady state and the largest eigenvalue $\lambda_1$ of the matrix $(-diag(\delta_i)+A)$, when the infection rate $\beta=1$ holds for all the links, the recovery rate of each node is assigned according to our strategy, i.e. $\delta_i=c_{2}k_i^\alpha$ and the average recovery rate $\langle\delta\rangle$ is given as shown in Fig. \ref{fig1}. In Fig. \ref{fig1a}, we can see that when the infection fraction $y_\infty=0$, the largest eigenvalue $\lambda_1$ is also close to $0$ which supports that $\lambda_1(-diag(\delta_i)+\beta*A)<0$ is the sufficient condition for an epidemic to die out. The trends of the two curves are consistent-the larger the eigenvalue $\lambda_1$ is, the larger the infection fraction $y_\infty$ is. However, their trends are not the same when the average recovery rate $\langle\delta\rangle$ decreases (i.e. the recovery resources are reduced) as shown in Fig. \ref{fig1b} and \ref{fig1c}. In other words, we cannot use the magnitude of $\lambda_1$ to indicate what extent the network is infected which suggests that we have to design recovery allocation strategies to minimize the fraction of infection.

\subsection{The influence of $\alpha$ or ($\alpha_{in},\alpha_{out}$)}
\label{sec3b}
We can also see in Fig. \ref{fig1}, compared to the homogeneous SIS model ($\alpha=0$), by applying our strategy to allocate the recovery rate, the infection fraction $y_\infty$ can be significantly reduced in a wide range of $\alpha$, and we will discuss the maximum possible improvement in detail in the next section. Moreover, the infection fraction is again increasing if the exponent $\alpha$ is larger than a certain value. That's to say, more recovery resources distributed to the high-degree nodes is not necessarily better in preventing infections; instead, too unbalanced distribution helps the virus. In the extreme case where $\alpha\rightarrow+\infty$, all the recovery rate is allocated to a single node, the one with the highest degree. In this case, all the other nodes have recovery rate 0. The network always gets fully infected except for the largest degree node, as long as the network is still connected after removing the highest degree node.

We investigate next the case of directed SF networks with directionality $\xi\approx1$ and a given indegree outdegree correlation $\rho$. Two scaling exponents $\alpha_{in}$ and $\alpha_{out}$ can be tuned to decide the recovery rate  $\delta_i=ck_{i,in}^{\alpha_{in}}k_{i,out}^{\alpha_{out}}$. We plot the infection fraction $y_\infty$ as a function of $\alpha_{in}$ and $\alpha_{out}$. In Fig. \ref{fig2a}, compared to the undirected networks with the same average recovery rate $\langle\delta\rangle = 3$, we find that most combinations of the two exponents $\alpha_{in}$ and $\alpha_{out}$ around $(0.5, 0.5)$, i.e. $\alpha_{in}\approx0.5,\alpha_{out}\approx0.5$, can lead to the die-out steady state. 
If we further decrease the average recovery rate $\langle\delta\rangle$, though the die-out steady state can not be reached any more, many combinations of $(\alpha_{in}, \alpha_{out})$ can reduce the infection fraction effectively compared to the homogeneous case, where $\alpha_{in}=\alpha_{out}=0$, as shown in Fig. \ref{fig2b}. Fig. \ref{fig2c} describes the case with a smaller average recovery rate $\langle\delta\rangle$ but larger indegree outdegree correlation ($\rho=0.5$), the same phenomenon can be observed.

    
\section{Optimal heterogeneous recovery allocations}
\label{sec4}
Most interestingly, we would like to find out our optimal strategy, i.e. the optimal exponent $\alpha_{opt}$ (or a combination $(\alpha_{in,opt}, \alpha_{out,opt})$) which leads to the minimum infection infraction $y_\infty$. Moreover, how much can our optimal strategy further reduce the fraction $y_\infty$ of infection compared to the homogeneous allocation of recovery rate?

\subsection{The optimal exponents $\alpha$ or $(\alpha_{in}, \alpha_{out})$}
\label{sec4a}
We list the optimal exponent $\alpha$ for undirected SF networks with different average recovery rates $\langle\delta\rangle$ in Table \ref{tab1}, and the optimal combination $(\alpha_{in,opt}, \alpha_{out,opt})$ of the two exponents $\alpha_{in}$ and $\alpha_{out}$ for directed SF networks with different average recovery rates $\langle\delta\rangle$ and indegree outdegree correlation $\rho$ in Table \ref{tab2}. For some combinations of $\langle\delta\rangle$ and $\rho$, like  $\langle\delta\rangle=3$ and $\rho=0$, more than one pair of ($\alpha_{in},\alpha_{out}$) can reduce the infection fraction $y_\infty$ to $0$, so we cannot list the optimal one. Note that, when indegree outdegree correlation $\rho=1$, the indegree and outdegree are the same for each node, so it is enough to list only one optimal exponent. We still confine ourselves to directed networks with directionality $\xi\approx1$. 

\begin{table}[!tbhp]
\renewcommand{\arraystretch}{1.3}
\caption{The optimal exponent $\alpha_{opt}$ for undirected SF networks with different average recovery rates $\langle\delta\rangle$}
\label{tab1}
\centering
\begin{tabular}{|c||c|c|c|c|c|c|}
\hline$\langle\delta\rangle$ & 1.5 & 1.75 & 2 & 2.25 & 2.5 & 2.75$\scriptsize{\sim}$4\\
\hline
$\alpha_{opt}$ & 0.3 & 0.4 & 0.7 & 0.8 & 0.9 & 1 \\
\hline
\end{tabular}
\end{table}

\begin{table}[!tbhp]
\renewcommand{\arraystretch}{1.3}
\caption{The optimal combinations ($\alpha_{in,opt}, \alpha_{out,opt}$) for directed SF networks with different $\langle\delta\rangle$ and $\rho$ }
\label{tab2}
\centering
\begin{tabular}{|c||p{1.3cm}<{\centering}|p{1.3cm}<{\centering}|p{1cm}<{\centering}|}\hline
\backslashbox{$\langle\delta\rangle$}{$\rho$} & 0 & 0.5 & 1\\
\hline
1.5  & ( 0.1, 0.6) & (-0.1, 0.6) & 0.1\\
\hline
1.75 & ( 0.2, 0.6) & ( 0.1, 0.6) & 0.3\\
\hline
2    & ( 0.4, 0.5) & ( 0.3, 0.6) & 0.5\\
\hline
2.25 & ( 0.5, 0.5) & ( 0.4, 0.5) & 0.7\\
\hline
2.5  & ( 0.5, 0.5) & ( 0.5, 0.5) & 0.8\\
\hline
2.75 &      -      & ( 0.5, 0.5) & 0.9\\
\hline
3    &      -      &       -      & 1\\
\hline
\end{tabular}
\end{table}

By observing Table \ref{tab1} and \ref{tab2}, we find that, with indegree outdegree correlation $\rho=1$, either the network is undirected or directed, the optimal exponent $\alpha_{opt}$ decreases as the average recovery rate $\langle\delta\rangle$ decreases. A positive $\alpha_{opt}$ implies that , we should distribute more recovery resources to the nodes with larger degrees. As the total recovery resources, $\langle\delta\rangle$ decrease, $\alpha_{opt}$ decreases, suggesting a relative homogeneous allocation of the recovery resources. When the recovery resources are large, $\alpha_{opt}$ is large, implying that a more heterogeneous distribution is most beneficial. Moreover, for directed networks with unequal indegree and outdegree, regardless of the average recovery rate $\langle\delta\rangle$, the optimal exponent $\alpha_{out,opt}$ for outdegree is always around $0.5$, though the optimal exponent $\alpha_{in,opt}$ for indegree declines as $\langle\delta\rangle$ decreases. In other words, the nodes with larger outdegree should always be equipped with faster recovery even if the total recovery resources are very limited, but not the nodes with larger indegree. That's reasonable because a larger outdegree means more chances to spread the virus and a fast recovery of such nodes is effective to prevent infections. 

\subsection{Optimal heterogeneous vs. homogeneous recovery rates}
Furthermore, we would like to understand how much our optimal strategy could further reduce the infection compared with classic homogeneous recovery rates allocation. We compare the infection fraction $y_\infty$ of the homogeneous ($\alpha=0$ or $\alpha_{in}=\alpha_{out}=0$) and our optimal heterogeneous ($\alpha=\alpha_{opt}$ or $\alpha_{in}=\alpha_{in,opt}, \alpha_{out}=\alpha_{out,opt}$) recovery rates allocation.
    
\begin{figure}[!tb]
\centering
\includegraphics[width=2.8in]{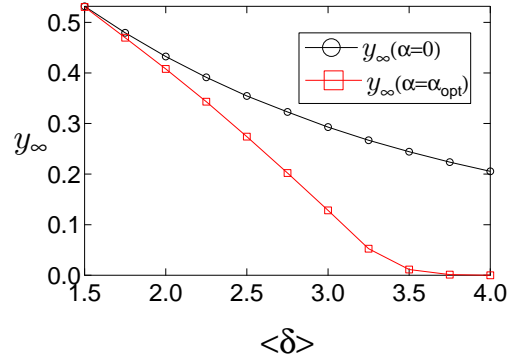}
\caption{The plot of the infection fraction $y_\infty$ as a function of the average recovery rate $\langle\delta\rangle$ for undirected SF networks under both homogeneous ($\alpha=0$) and optimal ($\alpha=\alpha_{opt}$) recovery rates allocation.}
\label{fig3}
\end{figure}

       \begin{figure*}[!tb]
          \centering
          \subfloat[$\rho=0$\label{fig4a}]{%
            \includegraphics[width=2.3in]{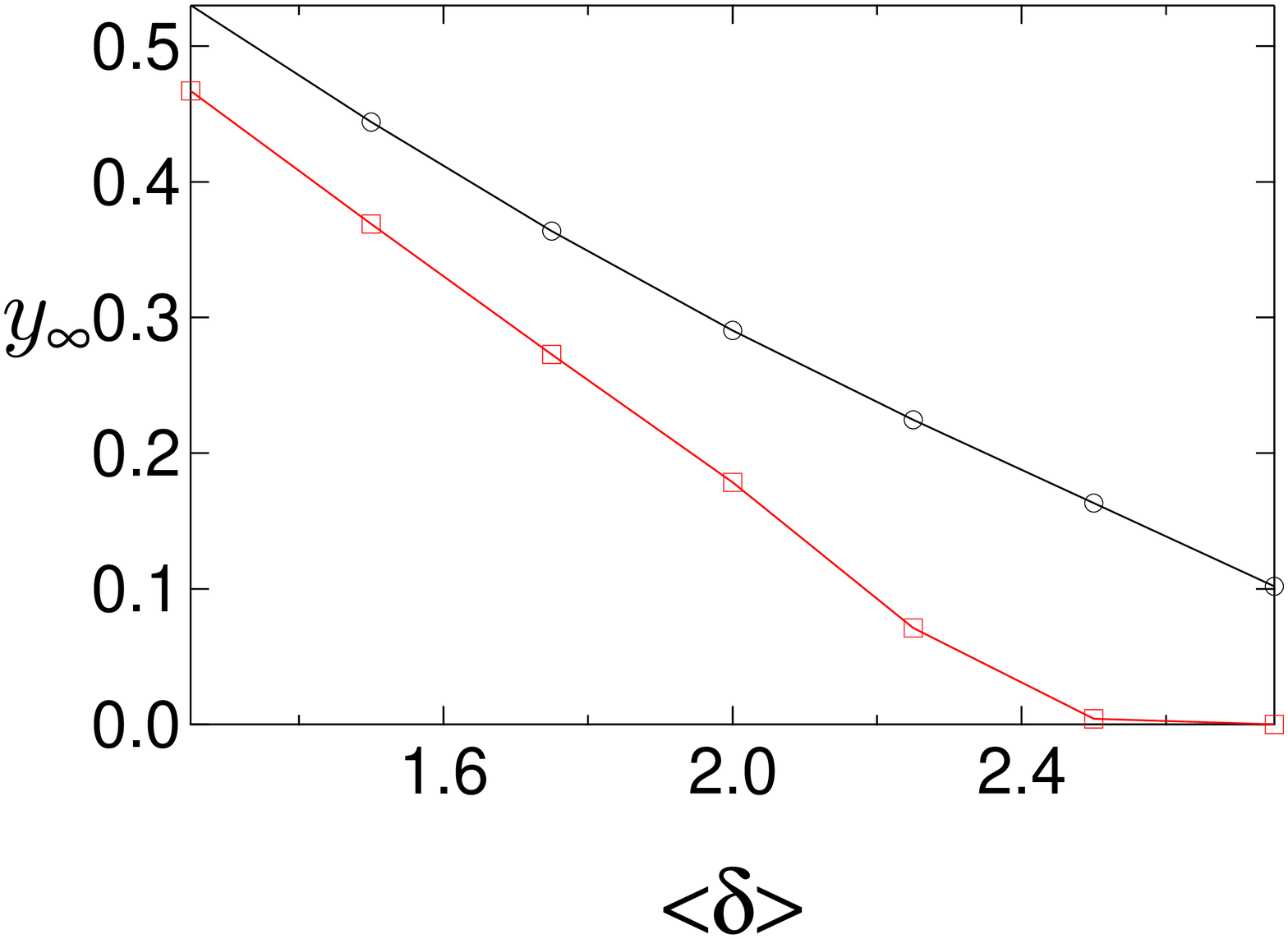}%
          }
          \hfill
          \subfloat[$\rho=0.5$\label{fig4b}]{%
            \includegraphics[width=2.3in]{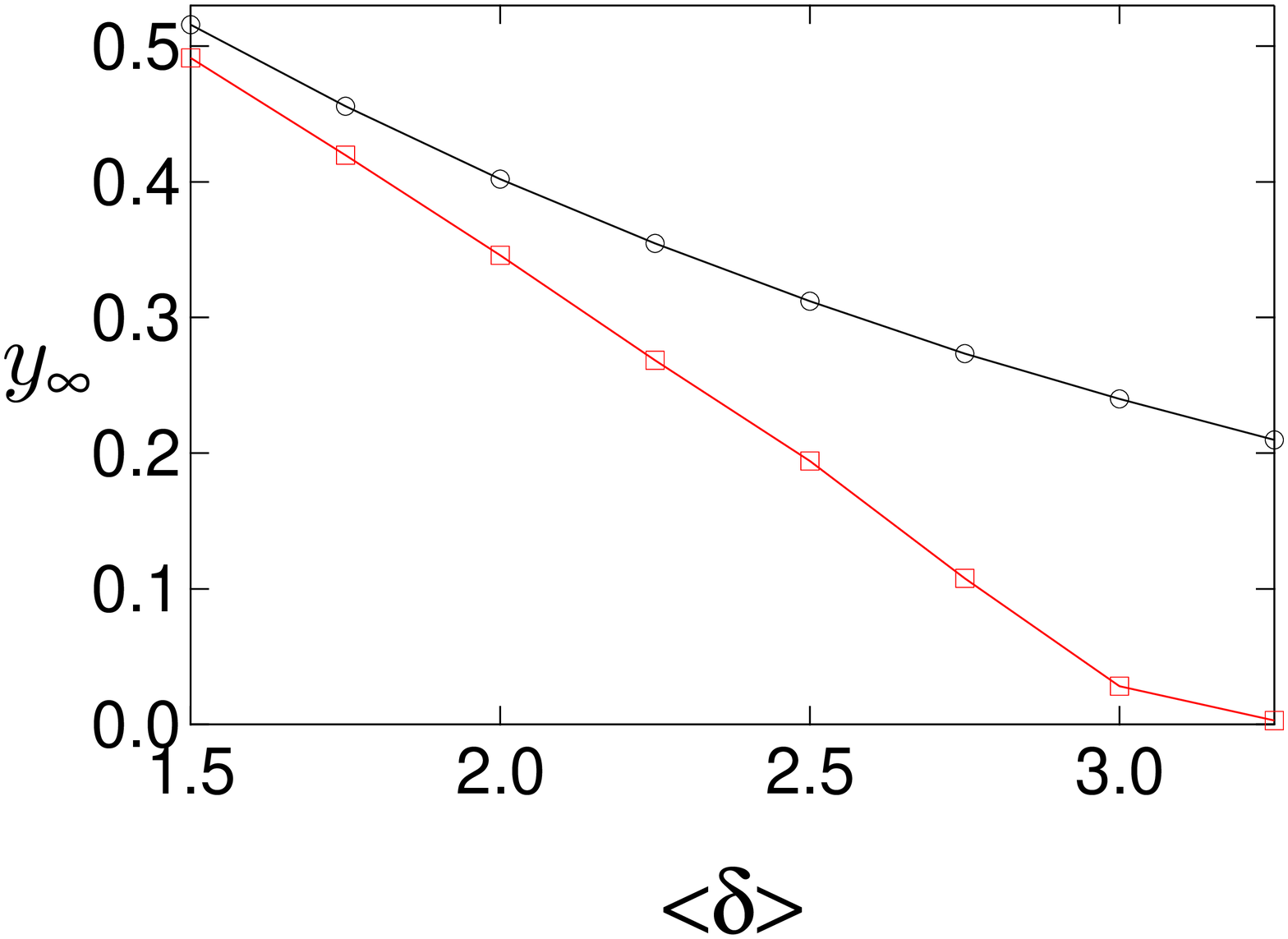}%
          }
          \hfill
                \subfloat[$\rho=1$\label{fig4c}]{%
                  \includegraphics[width=2.3in]{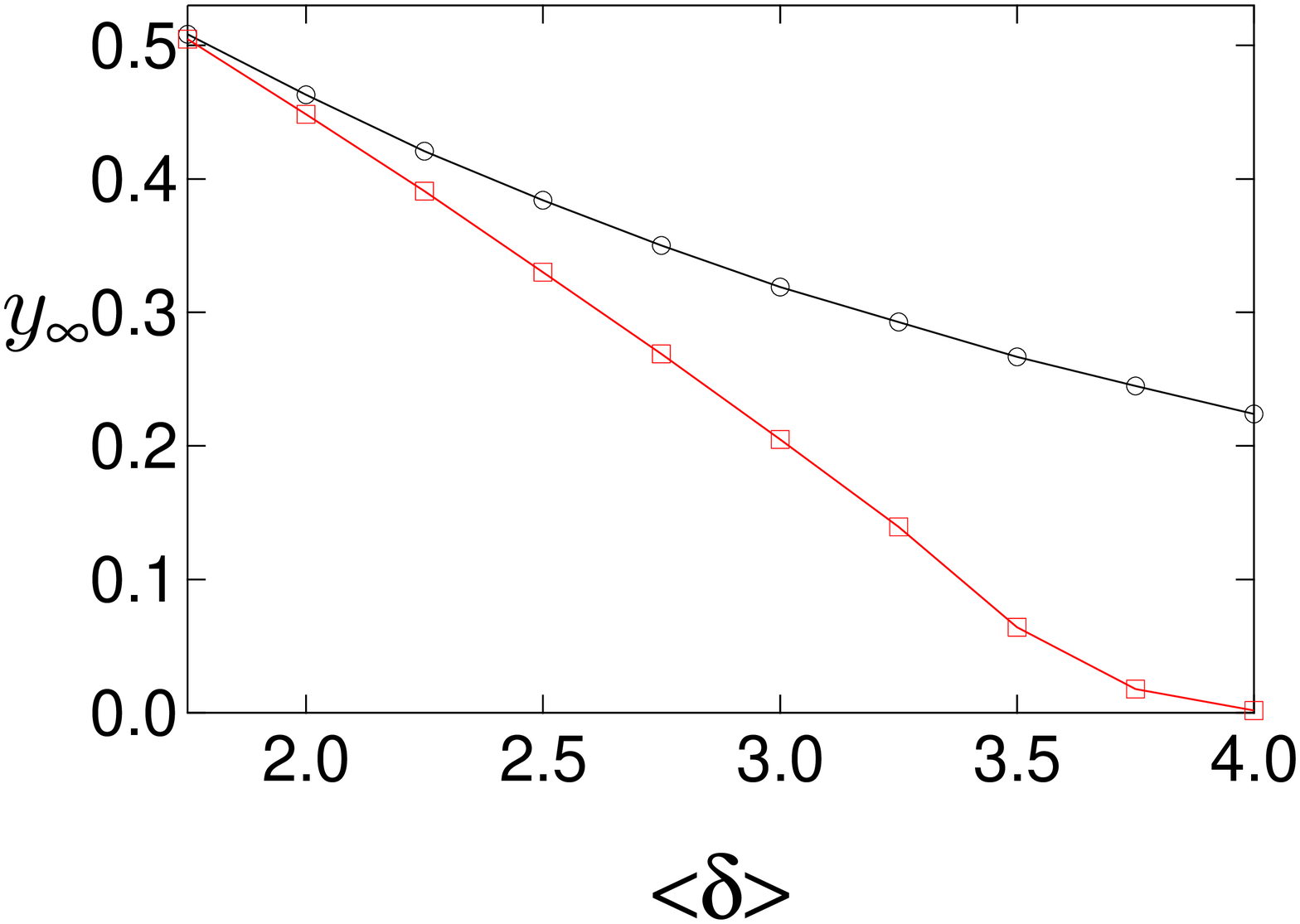}%
                }
          \caption{The infection fraction $y_\infty$ as a function of the average recovery rate $\langle\delta\rangle$ in directed SF networks with different indegree outdegree correlation $\rho$, under both homogeneous ($\textcolor{black}{\circ}$, $\alpha_{in}=\alpha_{out}=0$) and optimal ($\textcolor{red}{\square}$, $\alpha_{in}=\alpha_{in, opt}, \alpha_{out}=\alpha_{out, opt}$) recovery rates allocation.}
          \label{fig4}
        \end{figure*}
Fig. \ref{fig3} shows the results for undirected SF networks, and Fig. \ref{fig4} for the directed SF networks with different indegree outdegree correlation $\rho$. We find that, in general, our optimal heterogeneous strategy outperforms the homogeneous one and such outperformance becomes more evident when the average recovery rate $\langle\delta\rangle$ is not too small (not close to the infection rate $\beta=1$). Our optimal heterogeneous strategy could even reduce the outbreak epidemic state with the homogeneous strategy into a die out state when the average recovery rate $\langle\delta\rangle$ is close to $3$.  

Fig. \ref{fig4} shows that a smaller indegree outdegree correlation $\rho$ retards the virus spreading given the recovery rate allocation strategy, e.g. the homogeneous allocation. A smaller indegree outdegree correlation $\rho$ implies the larger difference between the indegree and outdegree of nodes. On one hand, a node with large outdegree may have a small indegree. Though such nodes have more chances to spread the virus, they are not likely to be infected. On the other hand, a node with a small outdegree may have a large indegree. Such nodes tend to be infected but do not help the virus to spread. Hence, a smaller indegree outdegree correlation $\rho$ prevents the spreading of the virus.

\begin{figure}[!tb]
\centering
\includegraphics[width=2.8in]{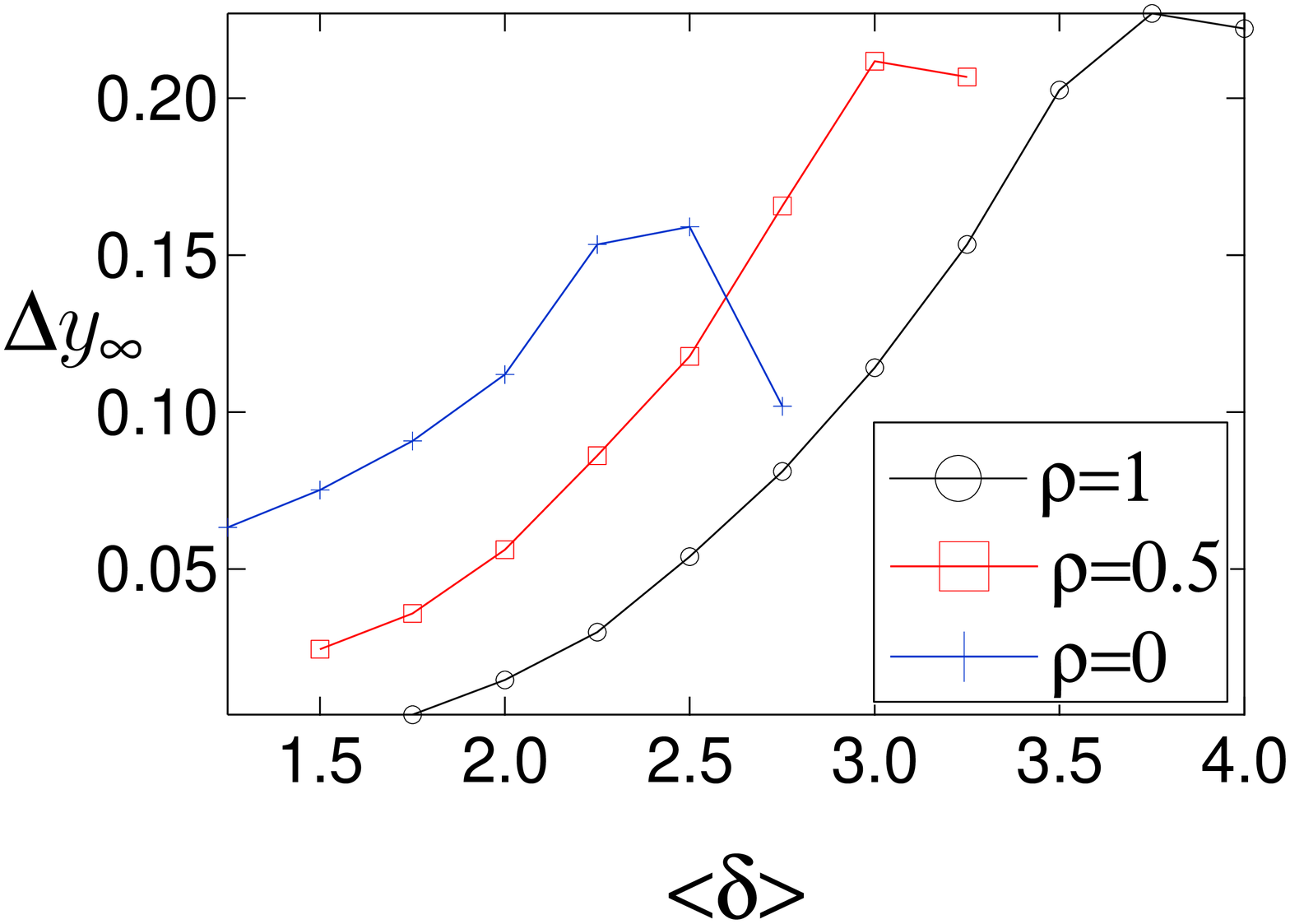}
\caption{The variation of the infection fraction $\Delta y_\infty$ as a function of the average recovery rate $\langle\delta\rangle$ for different values of the indegree outdegree correlation $\rho$.}
\label{fig5}
\end{figure}

We further explore the performance of our optimal heterogeneous recovery rates allocation strategy for different indegree outdegree correlation $\rho$. We calculate the further improvement in the infection fraction, $\Delta y_\infty$, as the difference between the infection fraction of the homogeneous ($y_{\infty,homo}$) and our optimal heterogeneous ($y_{\infty,opt}$) recovery rates allocation, i.e. $\Delta y_\infty = y_{\infty,homo}-y_{\infty,opt}$, given the same average recovery rate $\langle\delta\rangle$. In Fig. \ref{fig5}, for different indegree outdegree correlation $\rho$, we plot $\Delta y_\infty$ as a function of the average recovery rate $\langle\delta\rangle$ to show how much our optimal strategy can reduce the infection fraction, compared to the homogeneous SIS model. There seems to be a peak at $\langle\delta\rangle_c$ in each curve. That's mainly because as $\langle\delta\rangle$ increases, both $y_{\infty,homo}$ and $y_{\infty,opt}$ decrease to $0$, limiting the difference between the heterogeneous and homogeneous strategy. Before the peak, i.e. when $\delta<\langle\delta\rangle_c$, we find that the effect of our optimal strategy is more evident when the average recovery rate $\langle\delta\rangle$ is large and when the indegree outdegree correlation $\rho$ is small. When the total recovery resources, i.e. the average recovery rate is very small or very large, the epidemic will anyway spread out or die out respectively, which is independent of the recovery rate allocation strategy. Whereas in between these two extreme cases, i.e. with intermediate $\langle\delta\rangle$, we always see the outperformance of our heterogeneous strategy. Such outperformance appears in a range of $\langle\delta\rangle$ with smaller values when the indegree outdegree correlation $\rho$ is smaller, likely due to the fact that a smaller degree correlation contributes to the prevention of epidemic spreading. 

\section{Conclusion}
\label{sec5}
In this work, we address a new challenging question: how to allocate the limited recovery resources heterogeneously so that the fraction of infection can be minimized? We propose the heterogeneous recovery rates allocation strategy which allocates different recovery rates to different nodes. Our strategy is based on the degree of each node, which has the lowest computational complexity, $\delta_i=ck_{i,in}^{\alpha_{in}}k_{i,out}^{\alpha_{out}}$. We consider both undirected and directed networks, characterized by the directionality $\xi$ and the indegree outdegree correlation $\rho$.

Interestingly, our strategy via the optimal choice of the parameters $\alpha_{in}$ and $\alpha_{out}$, evidently outperforms the classic homogeneous allocation of recovery resources in general, especially when the given recovery resources are sufficient. The optimal choice of the parameter $\alpha_{in}$ and $\alpha_{out}$ depends on the indegree outdegree correlation $\rho$ and average recovery rate $\langle\delta\rangle$. We find that in undirected networks, when the average recovery rate $\langle\delta\rangle$ is large, $\alpha_{opt}$ is large, meaning that we should allocate the the recovery resources more heterogeneously. In directed networks, it seems that we should allocate more recovery resources to the high outdegree nodes, whereas the indegree has minor influence in determining the recovery rate as the average recovery rate $\langle\delta\rangle$ decreases.

Our degree based heterogeneous recovery rates allocation strategy illustrates the potential to more effectively reduce infection than the classic homogeneous allocation. However, this is just a start and hopefully it could inspire better heterogeneous strategies. It would be interesting to further consider the heterogeneous infection rate, motivated by real-world data set. The allocation of recovery rates in that case is far from well understood.


\section*{Acknowledgment}

This work has been partially supported by the European Commission within the framework of the CONGAS project
FP7-ICT-2011-8-317672 and the China Scholarship Council (CSC).




\begin{thebibliography}{10}
\providecommand{\url}[1]{#1}
\csname url@samestyle\endcsname
\providecommand{\newblock}{\relax}
\providecommand{\bibinfo}[2]{#2}
\providecommand{\BIBentrySTDinterwordspacing}{\spaceskip=0pt\relax}
\providecommand{\BIBentryALTinterwordstretchfactor}{4}
\providecommand{\BIBentryALTinterwordspacing}{\spaceskip=\fontdimen2\font plus
\BIBentryALTinterwordstretchfactor\fontdimen3\font minus
  \fontdimen4\font\relax}
\providecommand{\BIBforeignlanguage}[2]{{%
\expandafter\ifx\csname l@#1\endcsname\relax
\typeout{** WARNING: IEEEtran.bst: No hyphenation pattern has been}%
\typeout{** loaded for the language `#1'. Using the pattern for}%
\typeout{** the default language instead.}%
\else
\language=\csname l@#1\endcsname
\fi
#2}}
\providecommand{\BIBdecl}{\relax}
\BIBdecl

\bibitem{albert2000error}
R.~Albert, H.~Jeong, and A.-L. Barab{\'a}si, ``Error and attack tolerance of
  complex networks,'' \emph{Nature}, vol. 406, no. 6794, pp. 378--382, 2000.

\bibitem{kephart1991directed}
J.~O. Kephart and S.~R. White, ``Directed-graph epidemiological models of
  computer viruses,'' in \emph{Research in Security and Privacy, 1991.
  Proceedings., 1991 IEEE Computer Society Symposium on}.\hskip 1em plus 0.5em
  minus 0.4em\relax IEEE, 1991, pp. 343--359.

\bibitem{pastor2001epidemic}
R.~Pastor-Satorras and A.~Vespignani, ``Epidemic dynamics and endemic states in
  complex networks,'' \emph{Physical Review E}, vol.~63, no.~6, p. 066117,
  2001.

\bibitem{garetto2003modeling}
M.~Garetto, W.~Gong, and D.~Towsley, ``Modeling malware spreading dynamics,''
  in \emph{INFOCOM 2003. Twenty-Second Annual Joint Conference of the IEEE
  Computer and Communications. IEEE Societies}, vol.~3.\hskip 1em plus 0.5em
  minus 0.4em\relax IEEE, 2003, pp. 1869--1879.

\bibitem{ganesh2005effect}
A.~Ganesh, L.~Massouli{\'e}, and D.~Towsley, ``The effect of network topology
  on the spread of epidemics,'' in \emph{INFOCOM 2005. 24th Annual Joint
  Conference of the IEEE Computer and Communications Societies. Proceedings
  IEEE}, vol.~2.\hskip 1em plus 0.5em minus 0.4em\relax IEEE, 2005, pp.
  1455--1466.

\bibitem{canright2006spreading}
G.~S. Canright and K.~Eng{\o}-Monsen, ``Spreading on networks: a topographic
  view,'' \emph{Complexus}, vol.~3, no. 1-3, pp. 131--146, 2006.

\bibitem{anderson1991infectious}
R.~M. Anderson and R.~M. May, \emph{Infectious diseases of humans}.\hskip 1em
  plus 0.5em minus 0.4em\relax Oxford university press Oxford, 1991, vol.~1.

\bibitem{bailey1975mathematical}
N.~T. Bailey \emph{et~al.}, \emph{The mathematical theory of infectious
  diseases and its applications}.\hskip 1em plus 0.5em minus 0.4em\relax
  Charles Griffin \& Company Ltd, 5a Crendon Street, High Wycombe, Bucks HP13
  6LE., 1975.

\bibitem{daley2001epidemic}
D.~J. Daley, J.~Gani, and J.~M. Gani, \emph{Epidemic modelling: an
  introduction}.\hskip 1em plus 0.5em minus 0.4em\relax Cambridge University
  Press, 2001, vol.~15.

\bibitem{van2011n}
P.~Van~Mieghem, ``The {N}-intertwined {SIS} epidemic network model,''
  \emph{Computing}, vol.~93, no. 2-4, pp. 147--169, 2011.

\bibitem{cator2013susceptible}
E.~Cator and P.~Van~Mieghem, ``Susceptible-infected-susceptible epidemics on
  the complete graph and the star graph: Exact analysis,'' \emph{Physical
  Review E}, vol.~87, no.~1, p. 012811, 2013.

\bibitem{li2012susceptible}
C.~Li, R.~van~de Bovenkamp, and P.~Van~Mieghem,
  ``Susceptible-infected-susceptible model: A comparison of n-intertwined and
  heterogeneous mean-field approximations,'' \emph{Physical Review E}, vol.~86,
  no.~2, p. 026116, 2012.

\bibitem{allen1994some}
L.~J. Allen, ``Some discrete-time {SI}, {SIR}, and {SIS} epidemic models,''
  \emph{Mathematical biosciences}, vol. 124, no.~1, pp. 83--105, 1994.

\bibitem{zhou2003stability}
Y.~Zhou and H.~Liu, ``Stability of periodic solutions for an sis model with
  pulse vaccination,'' \emph{Mathematical and Computer Modelling}, vol.~38,
  no.~3, pp. 299--308, 2003.

\bibitem{boccara1993critical}
N.~Boccara and K.~Cheong, ``Critical behaviour of a probabilistic automata
  network {SIS} model for the spread of an infectious disease in a population
  of moving individuals,'' \emph{Journal of Physics A: Mathematical and
  General}, vol.~26, no.~15, p. 3707, 1993.

\bibitem{shi2008sis}
H.~Shi, Z.~Duan, and G.~Chen, ``An {SIS} model with infective medium on complex
  networks,'' \emph{Physica A: Statistical Mechanics and its Applications},
  vol. 387, no.~8, pp. 2133--2144, 2008.

\bibitem{korobeinikov2002lyapunov}
A.~Korobeinikov and G.~C. Wake, ``Lyapunov functions and global stability for
  sir, sirs, and sis epidemiological models,'' \emph{Applied Mathematics
  Letters}, vol.~15, no.~8, pp. 955--960, 2002.

\bibitem{ferreira2012epidemic}
S.~C. Ferreira, C.~Castellano, and R.~Pastor-Satorras, ``Epidemic thresholds of
  the susceptible-infected-susceptible model on networks: A comparison of
  numerical and theoretical results,'' \emph{Physical Review E}, vol.~86,
  no.~4, p. 041125, 2012.

\bibitem{caldarelli2000fractal}
G.~Caldarelli, R.~Marchetti, and L.~Pietronero, ``The fractal properties of
  internet,'' \emph{EPL (Europhysics Letters)}, vol.~52, no.~4, p. 386, 2000.

\bibitem{albert1999internet}
R.~Albert, H.~Jeong, and A.-L. Barab{\'a}si, ``Internet: Diameter of the
  world-wide web,'' \emph{Nature}, vol. 401, no. 6749, pp. 130--131, 1999.

\bibitem{faloutsos1999power}
M.~Faloutsos, P.~Faloutsos, and C.~Faloutsos, ``On power-law relationships of
  the internet topology,'' in \emph{ACM SIGCOMM Computer Communication Review},
  vol.~29, no.~4.\hskip 1em plus 0.5em minus 0.4em\relax ACM, 1999, pp.
  251--262.

\bibitem{siganos2003power}
G.~Siganos, M.~Faloutsos, P.~Faloutsos, and C.~Faloutsos, ``Power laws and the
  as-level internet topology,'' \emph{IEEE/ACM Transactions on Networking
  (TON)}, vol.~11, no.~4, pp. 514--524, 2003.

\bibitem{chen2002origin}
Q.~Chen, H.~Chang, R.~Govindan, and S.~Jamin, ``The origin of power laws in
  internet topologies revisited,'' in \emph{INFOCOM 2002. Twenty-First Annual
  Joint Conference of the IEEE Computer and Communications Societies.
  Proceedings. IEEE}, vol.~2.\hskip 1em plus 0.5em minus 0.4em\relax IEEE,
  2002, pp. 608--617.

\bibitem{ebel2002scale}
H.~Ebel, L.-I. Mielsch, and S.~Bornholdt, ``Scale-free topology of e-mail
  networks,'' \emph{Physical Review E}, vol.~66, no.~3, p. 035103, 2002.

\bibitem{barabasi2000scale}
A.-L. Barab{\'a}si, R.~Albert, and H.~Jeong, ``Scale-free characteristics of
  random networks: the topology of the world-wide web,'' \emph{Physica A:
  Statistical Mechanics and its Applications}, vol. 281, no.~1, pp. 69--77,
  2000.

\bibitem{JCN2013_Trajanovski}
S.~Trajanovski, J.~Mart\'{i}n-Hern\'{a}ndez, W.~Winterbach, and P.~{Van
  Mieghem}, ``Robustness envelopes of networks,'' \emph{Journal of Complex
  Networks}, 2013.

\bibitem{huang2013robustness}
X.~Huang, S.~Shao, H.~Wang, S.~V. Buldyrev, H.~E. Stanley, and S.~Havlin, ``The
  robustness of interdependent clustered networks,'' \emph{EPL (Europhysics
  Letters)}, vol. 101, no.~1, p. 18002, 2013.

\bibitem{li2012degree}
C.~Li, H.~Wang, and P.~Van~Mieghem, ``Degree and principal eigenvectors in
  complex networks,'' in \emph{NETWORKING 2012}.\hskip 1em plus 0.5em minus
  0.4em\relax Springer, 2012, pp. 149--160.

\bibitem{li2011correlation}
C.~Li, H.~Wang, W.~De~Haan, C.~Stam, and P.~Van~Mieghem, ``The correlation of
  metrics in complex networks with applications in functional brain networks,''
  \emph{Journal of Statistical Mechanics: Theory and Experiment}, vol. 2011,
  no.~11, p. P11018, 2011.

\bibitem{cohen2003efficient}
R.~Cohen, S.~Havlin, and D.~Ben-Avraham, ``Efficient immunization strategies
  for computer networks and populations,'' \emph{Physical review letters},
  vol.~91, no.~24, p. 247901, 2003.

\bibitem{madar2004immunization}
N.~Madar, T.~Kalisky, R.~Cohen, D.~ben Avraham, and S.~Havlin, ``Immunization
  and epidemic dynamics in complex networks,'' \emph{The European Physical
  Journal B-Condensed Matter and Complex Systems}, vol.~38, no.~2, pp.
  269--276, 2004.

\bibitem{pastor2002immunization}
R.~Pastor-Satorras and A.~Vespignani, ``Immunization of complex networks,''
  \emph{Physical Review E}, vol.~65, no.~3, p. 036104, 2002.

\bibitem{gomez2006immunization}
J.~G{\'o}mez-Gardenes, P.~Echenique, and Y.~Moreno, ``Immunization of real
  complex communication networks,'' \emph{The European Physical Journal
  B-Condensed Matter and Complex Systems}, vol.~49, no.~2, pp. 259--264, 2006.

\bibitem{bai2007immunization}
W.-J. Bai, T.~Zhou, and B.-H. Wang, ``Immunization of susceptible--infected
  model on scale-free networks,'' \emph{Physica A: Statistical Mechanics and
  its Applications}, vol. 384, no.~2, pp. 656--662, 2007.

\bibitem{fu2008epidemic}
X.~Fu, M.~Small, D.~M. Walker, and H.~Zhang, ``Epidemic dynamics on scale-free
  networks with piecewise linear infectivity and immunization,'' \emph{Physical
  Review E}, vol.~77, no.~3, p. 036113, 2008.

\bibitem{buono2013slow}
C.~Buono, F.~Vazquez, P.~Macri, and L.~Braunstein, ``Slow epidemic extinction
  in populations with heterogeneous infection rates,'' \emph{Physical Review
  E}, vol.~88, no.~2, p. 022813, 2013.

\bibitem{preciado2013optimal}
V.~M. Preciado, M.~Zargham, C.~Enyioha, A.~Jadbabaie, and G.~Pappas, ``Optimal
  vaccine allocation to control epidemic outbreaks in arbitrary networks,'' in
  \emph{Decision and Control (CDC), 2013 IEEE 52nd Annual Conference on}.\hskip
  1em plus 0.5em minus 0.4em\relax IEEE, 2013, pp. 7486--7491.

\bibitem{li2013epidemic}
C.~Li, H.~Wang, and P.~Van~Mieghem, ``Epidemic threshold in directed
  networks,'' \emph{Physical Review E}, vol.~88, no.~6, p. 062802, 2013.

\bibitem{qu2014non}
B.~Qu, Q.~Li, S.~Havlin, H.~E. Stanley, and H.~Wang, ``Non-consensus opinion
  model on directed networks,'' \emph{arXiv preprint arXiv:1404.7318}, 2014.

\bibitem{PhysRevLett.85.4626}
\BIBentryALTinterwordspacing
R.~Cohen, K.~Erez, D.~ben Avraham, and S.~Havlin, ``Resilience of the internet
  to random breakdowns,'' \emph{Phys. Rev. Lett.}, vol.~85, pp. 4626--4628, Nov
  2000. [Online]. Available:
  \url{http://link.aps.org/doi/10.1103/PhysRevLett.85.4626}
\BIBentrySTDinterwordspacing

\bibitem{PhysRevE.84.016101}
\BIBentryALTinterwordspacing
P.~Van~Mieghem, D.~Stevanovi\ifmmode~\acute{c}\else \'{c}\fi{}, F.~Kuipers,
  C.~Li, R.~van~de Bovenkamp, D.~Liu, and H.~Wang, ``Decreasing the spectral
  radius of a graph by link removals,'' \emph{Phys. Rev. E}, vol.~84, p.
  016101, Jul 2011. [Online]. Available:
  \url{http://link.aps.org/doi/10.1103/PhysRevE.84.016101}
\BIBentrySTDinterwordspacing

\bibitem{li2012bounds}
C.~Li, H.~Wang, and P.~Van~Mieghem, ``Bounds for the spectral radius of a graph
  when nodes are removed,'' \emph{Linear Algebra and its Applications}, vol.
  437, no.~1, pp. 319--323, 2012.

\bibitem{wang2013effect}
H.~Wang, Q.~Li, G.~D扐gostino, S.~Havlin, H.~E. Stanley, and P.~Van~Mieghem,
  ``Effect of the interconnected network structure on the epidemic threshold,''
  \emph{Physical Review E}, vol.~88, no.~2, p. 022801, 2013.

\end{thebibliography}
%

\end{document}